\documentclass[preprint,
showpacs,preprintnumbers,amsmath,amssymb]{revtex4}

\usepackage{graphicx}% Include figure files
\usepackage{dcolumn}% Align table columns on decimal point
\usepackage{bm,morefloats,color}% bold math

\usepackage{amssymb,bm}

\def\Eq#1{\begin{equation} #1 \end{equation}}
\def\Eqr#1{\begin{eqnarray} #1 \end{eqnarray}}
\def\Eqrsubl#1#2{\begin{subequations}\label{#1}\Eqr{#2}\end{subequations}}

\newcommand{\nn}{\nonumber}
\newcommand{\pd}{\partial}

\newcommand{\bea}{\begin{eqnarray}}
\newcommand{\eea}{\end{eqnarray}}

\def\Xsp{{\rm X}}
\def\bXsp{\bar{{\rm X}}}
\def\Ysp{{\rm Y}}
\def\Zsp{{\rm Z}}
\def\X5sp{{\rm X}_5}
\def\Y3sp{{\rm Y}_3}
\def\Z3sp{{\rm Z}_3}

\def\lap{{\triangle}}
\def\e{{\rm e}}

\begin{document}

%\preprint{YITP-11-102}

\title{\bf Warped spherical
compactifications in the gravity theory}

\author{Masato Minamitsuji}
\affiliation{
Yukawa Institute for Theoretical Physics
Kyoto University, Kyoto 606-8502, Japan.
}%

\author{Kunihito Uzawa}
\affiliation{
Department of Physics,
School of Science and Technology,
Kwansei Gakuin University, Sanda, Hyogo 669-1337, Japan.
}%

\date{\today}% It is always \today, today,

\begin{abstract}
We present new exact solutions 
of the warped spherical compactifications 
in the higher-dimensional gravitational theory 
coupled to scalar and several form field strengths.
We find two classes of solutions.  
One has a de Sitter spacetime with a static warp factor.  
The other gives an accelerating universe 
in the non-Einstein conformal frame 
with a time-dependent warp factor.
\end{abstract}

\pacs{04.20.-q, 04.20.Jb}% PACS, the Physics and Astronomy
\maketitle

%======================================%
%<<<<<<<<<<<<< SECTION 1 >>>>>>>>>>>>>>%
%======================================%

%T1>Introduction
\section{Introduction}
 \label{sec:introduction}
Recently, many models of warped compactifications 
have been studied in 
the higher-dimensional gravitational theory.
These models have shed light on many different aspects of cosmology,
phenomenology and black hole physics.  
The purpose of the present paper is 
to obtain new classes of warped solutions and 
apply them to study of the  
four-dimensional cosmological dynamics,
in particular a realization of 
a de Sitter or an accelerating universe.
A motivation for studying  
a de Sitter or an accelerating universe
is that it may give us a fairly direct explanation
of the inflationary universe 
and/or the current acceleration of the universe 
from the purely gravitational point of view 
\cite{Tsujikawa:2010sc}.  
So far, most works devoted to
compactifications to de Sitter or accelerating universes
have been made in the context of the
four-dimensional effective theory. 
However, straightforward derivations of such solutions  
in the higher-dimensional gravity have not been 
extensively explored so much in terms of 
finding the exact solutions of the field equations.
In particular, we will present warped cosmological solutions 
in which the internal space geometry is given by 
a product of several spheres. 
These are similar to the solutions discussed in \cite{Minamitsuji:2011xs}. 
Although we consider a simple class of the ansatz for fields, 
this provides a good illustration to find the warped cosmological solution.

A warped spherical compactification has been  
used in \cite{Minamitsuji:2011xs}.
The basic idea was  
to assume that the internal geometry is given 
by a product of S$^1$ and several other compact spaces,
which are spheres in our case,
and also that the warp factor depends only on the 
coordinate of S$^1$.
Along such a way,
with a more general class of the space
as S${}^1\times \prod_I$S${}^{L_I}$, 
we will study the compactification to a
maximally symmetric spacetime,
i.e., a de Sitter or an anti- de Sitter (AdS) universe. 
Because of the maximal symmetry of an external spacetime,
it can be embedded
into a warped spacetime 
where the warp factor and the internal space 
are independent of time.
In fact, the model we will focus on has been studied, 
from a rather different vantage
point 
in \cite{Neupane:2009jn, Neupane:2010is, Neupane:2010ya}. 
A number of important features were pointed out in \cite{Minamitsuji:2011xs}, 
including not only the gravitational field
but also matter fields. 
Such a model has also been investigated recently 
in the works \cite{Minamitsuji:2011gn, Minamitsuji:2011gp}
which have overlap with our present work,
but has still become interesting in light of 
other type of compactifications. 

 We also discuss an embedding of
 a more general expanding
Friedmann-Lema\^itre-Robertson-Walker (FLRW) universe
whose scale factor is given by a power-law
function of the cosmic time into the warped 
compactification on S${}^1\times \prod_I$S${}^{L_I}$.
Because of a less symmetry of an external geometry, 
it may be possible 
only into a higher-dimensional spacetime
where the warp factor and the internal space 
become time-dependent.
We will investigate
whether and how 
cosmic expansion
can be incorporated 
with a time-dependence of the warp and the internal space,
and,
if there are solutions,
whether 
they 
could realize  
accelerating universes.
Such  
solutions with 
time-dependence 
in the warp factor 
have been less investigated,
and 
exceptional cases 
are the time-dependent brane solutions
in the supergravity theories
\cite{Lu:1996jk, Behrndt:2003cx, 
Gibbons:2005rt, Chen:2005jp, Kodama:2005fz, Kodama:2005cz, Kodama:2006ay, 
Binetruy:2007tu, Binetruy:2008ev, Maeda:2009tq, Maeda:2009zi, Gibbons:2009dr, 
Maeda:2009ds, Uzawa:2010zza, Maeda:2010yk, Maeda:2010ja, Minamitsuji:2010fp, 
Maeda:2010aj, Minamitsuji:2010kb, Minamitsuji:2010uz, Uzawa:2010zz, 
Maeda:2011sh, Minamitsuji:2011jt, Maeda:2012xb, Minamitsuji:2012if},
which have direct connections with  
D-branes and M-branes in superstring and M-theory, respectively.
Since these dynamical brane solutions  
provide accelerated expansion of the four-dimensional universe
in the non-Einstein conformal frame, we will explore new cosmological 
solutions for the warped spacetime. 
We present an ansatz for a warped spacetime 
in the $D$-dimensional theory that 
realizes the four-dimensional de Sitter or AdS universe, 
and then we solve the higher-dimensional field equations.  

We start by describing warped de Sitter solutions 
in section \ref{sec:solution}. In section \ref{sec:s}, 
we discuss a time-dependent 
warped compactification with an infinite volume of the internal space. 
We summarize our results in section \ref{sec:Discussions}.

%======================================%
%<<<<<<<<<<<<< SECTION 2 >>>>>>>>>>>>>>%
%======================================%
%T1>Warped de Sitter solutions
\section{Warped de Sitter solutions}
\label{sec:solution}
In this section,
we focus on an embedding of 
a maximally symmetric,
a de Sitter or AdS,
universe into
a warped spacetime
where the warp factor and the internal space are static.

We consider a gravitational theory with the metric $g_{MN}$,
the scalar field $\phi$ 
and the antisymmetric tensor field of rank $p_I$.
The action which we consider is given by
\Eq{
S=\frac{1}{2\kappa^2}\int 
\left[R\ast{\bf 1}
-\frac{1}{2}d\phi\wedge\ast d\phi-\sum_I\frac{1}{2\cdot p_I!}
\e^{\varepsilon c_I\phi}
F_{(p_I)}\wedge\ast F_{(p_I)}\right],
\label{0:action:Eq}
}
where $\kappa^2$ is the $D$-dimensional gravitational constant, 
and $\ast$ is the Hodge operator in $D$-dimensional spacetime, 
$F_{(p_I)}$ is the $p_I$-form field strength,
and $c_I$ is a constant, and $\varepsilon=\pm 1$. 

The $D$-dimensional action (\ref{0:action:Eq}) gives the 
field equations: 
\Eqrsubl{0:field equations:Eq}{
&&\hspace{-0.8cm}R_{MN}=\frac{1}{2}\pd_M\phi\pd_N\phi
+\sum_I\frac{\e^{\varepsilon c_I\phi}}{2\cdot p_{I}!}
\left[p_I\left(F_{(p_I)}\right)^2_{MN}
-\frac{p_I-1}{D-2}g_{MN}F_{(p_I)}^2\right],
   \label{0:Einstein:Eq}\\
&&\hspace{-0.8cm}
d\left[\e^{\varepsilon c_I\phi}\,\ast F_{(p_I)}\right]=0\,,
    \label{0:gauge:Eq}\\
&&\hspace{-0.8cm}d\ast d\phi
-\sum_I\frac{\varepsilon c_I}{2\cdot p_I!}\e^{\varepsilon c_I\phi}F_{(p_I)}
\wedge\ast F_{(p_I)}=0\,,
   \label{0:scalar:Eq}
}
where $\left(F_{(p_I)}\right)^2_{MN}
= F_{MA_1\cdots A_{p_I-1}}
{F_N}^{A_1\cdots A_{p_I-1}}$.
In this paper, 
the derivation of the solutions
completely follows
the method developed in the previous works
devoted to
finding the solutions of the 
warped de Sitter compactifications
\cite{Minamitsuji:2011xs, Neupane:2010ya, Minamitsuji:2011gn, 
Minamitsuji:2011gp}.
To solve the field equations, 
we look for solutions whose spacetime metric has the form
\Eq{
ds^2=\e^{2A(y)}\left[q_{\mu\nu}(\Xsp)dx^{\mu}dx^{\nu}+dy^2(\Ysp)
+\sum_Iu_{a_Ib_I}(\Zsp_I)dz^{a_I}dz^{b_I}\right],
 \label{0:metric:Eq}
}
where $q_{\mu\nu}(\Xsp)$ is %a 
the $n$-dimensional metric which
depends only on the $n$-dimensional coordinates $x^{\mu}$,
and $u_{a_Ib_I}(\Zsp_I)$ is the $L_I$-dimensional metric which
depends only on the $L_I$-dimensional coordinates $z^{a_I}$, the function 
$A(y)$ depends only on the coordinate $y$, respectively. 
Hence, the warp factor is limited to the function $A$.
The number of dimensions 
is related as
$D=n+\sum_I L_I+1$.
Although we consider the particular class of solutions,
they provide us a direct route to warped de Sitter solutions.

Concerning the gauge fields, we adopt the following assumptions
\Eqrsubl{0:fields:Eq}{
F_{(n)}&=&f_n\,\Omega(\Xsp)\,,
  \label{0:ansatz of gauge1:Eq}\\
F_{(L_I)}&=&\ell_I\,\omega_I\,,
  \label{0:ansatz of gauge2:Eq}\\
F_{(0)}&=&m\,,
  \label{0:ansatz of gauge3:Eq}
}
where $f_n$, $\ell_I$ and $m$ 
are constants, and $c_0$ is the coupling constant for $F_{(0)}$, 
and $\Omega(\Xsp)$ and $\omega_I$ 
denote the volume $n$-form and the $L_I$-form for each $\Zsp_I$, respectively. 
Under the $D$-dimensional metric and form fields given above, 
we first reduce the Einstein equations other than the gauge and 
scalar field equations to a simple set of equations
\Eqrsubl{g:cEinstein0:Eq}{
&&\hspace{-1.1cm}
R_{\mu\nu}(\Xsp)-\frac{1}{2}\pd_{\mu}\phi\pd_{\nu}\phi-\left[U
-\frac{1}{2(D-2)}\left\{(D-n-3)f_n^2\,\e^{-2(n-1)A
+\varepsilon c_n\phi}\right.\right.\nn\\
&&\left.\left.~~+\sum_I(L_I-1)\ell_I^2\,\e^{-2(L_I-1)A
+\varepsilon c_I\phi}-m^2\,\e^{2A+\varepsilon c_0\phi}\right\}
\right]q_{\mu\nu}(\Xsp)
=0,
 \label{0:cEinstein-mn0:Eq}\\
&&\hspace{-1.1cm}
-(D-1)\pd_y^2A-\frac{1}{2}\left(\pd_y\phi\right)^2
-\frac{1}{2(D-2)}\left[(n-1)f_n^2\,\e^{-2(n-1)A
+\varepsilon c_n\phi}\right.\nn\\
&&\left.~~-\sum_I(L_I-1)\ell_I^2\,\e^{-2(L_I-1)A
+\varepsilon c_I\phi}+m^2\,\e^{2A+\varepsilon c_0\phi}\right]
=0,
  \label{0:cEinstein-yy0:Eq}\\
&&\hspace{-1.1cm}R_{a_Ib_I}(\Zsp_I)-\frac{1}{2}\pd_{a_I}\phi\pd_{b_I}\phi
-\left[U+\frac{1}{2(D-2)}\left\{(n-1)f_n^2\,\e^{-2(n-1)A
+\varepsilon c_n\phi}\right.\right.\nn\\
&&\left.\left.~~+\sum_I(D-L_I-3)\ell_I^2\,\e^{-2(L_I-1)A
+\varepsilon c_I\phi}+m^2\,\e^{2A+\varepsilon c_0\phi}\right\}
\right]u_{a_Ib_I}(\Zsp_I)=0,
 \label{0:cEinstein-ab0:Eq}
}
where $R_{\mu\nu}(\Xsp)$, $R_{a_Ib_I}(\Zsp_I)$ are the Ricci tensors
of the metrics $q_{\mu\nu}(\Xsp)$, $u_{a_Ib_I}(\Zsp_I)$, and 
$U$ is defined by 
\Eq{
U=\pd_y^2A+(D-2)\left(\pd_yA\right)^2\,.
  \label{0:U:Eq}
}
Next we consider the scalar field. 
We require that the scalar field satisfies the condition
\Eq{
2A+\varepsilon c_0\phi=0\,,~~~~-2(n-1)A
+\varepsilon c_n\phi=0\,,~~~~-2(L_I-1)A
+\varepsilon c_I\phi=0.
  \label{0:ansatz of scalar:Eq}
}
With these expressions, we immediately see that the scalar field 
$\phi$ depends on the linear function of the warp factor and the 
exponents on the exponential functions in the Einstein 
equations \eqref{g:cEinstein0:Eq} vanish separately. 
Hence, \eqref{0:ansatz of scalar:Eq} is equivalent to
\Eq{
\phi=-\frac{2A}{\varepsilon c_0}\,,~~~~
c_{n}=-\left(n-1\right)%\varepsilon 
c_0\,,~~~~
c_{I}=-\left(L_I-1\right)%\varepsilon 
c_0\,,
   \label{0:coupling:Eq}
}
where $c_{n}$ and $c_{I}$ are 
the coupling constants for $F_{(n)}$ and $F_{(I)}$, respectively.  
Under the assumptions (\ref{0:fields:Eq}) and (\ref{0:coupling:Eq}), 
the Bianchi identities and the equations of 
motion for the gauge fields are automatically satisfied.

We should not choose $m=0$ because of the ansatz of the scalar field 
(\ref{0:ansatz of scalar:Eq}). 
Thus, we focus on the case of $m\neq 0$ 
and come back to the case of $m=0$ later.
The choice of the ansatz for 
the metric and matter fields 
used to 
solve field equations 
is essential, and a particular coupling constant is 
preferred in order to 
make both  
the equations of motion and 
the form fields for the internal spaces including the 0-form
simple. 

Substituting Eqs.~(\ref{0:metric:Eq}), (\ref{0:fields:Eq}) 
and (\ref{0:coupling:Eq}) into
Eq.~(\ref{0:scalar:Eq}), the scalar field equation gives
\Eqr{
U+\frac{c_0^2}{4}\left(
K+m^2\right)=0\,,
   \label{0:scalar eq:Eq}
}
where $K$ is defined as
\Eq{
K= (n-1)f_n^2-\sum_I(L_I-1)f_I^2\,.
  \label{0:K:Eq}
} 
Using the scalar field equation (\ref{0:scalar eq:Eq}), 
the Einstein equations (\ref{g:cEinstein0:Eq}) are rewritten as
\Eqrsubl{g:cEinstein:Eq}{
&&\hspace{-1.1cm}
R_{\mu\nu}(\Xsp)-\frac{1}{2}\left(2U-f_n^2+\frac{K+m^2}{D-2}\right)
q_{\mu\nu}(\Xsp)
=0,
 \label{0:cEinstein-mn:Eq}\\
&&\hspace{-1.1cm}
-(D-2)\pd_y^2A+\left(D-2-\frac{%1
2}{c_0^2}\right)\left(\pd_yA\right)^2
-\frac{1}{2}\left(2U+\frac{K+m^2}{D-2}\right)
=0,
  \label{0:cEinstein-yy:Eq}\\
&&\hspace{-1.1cm}R_{a_Ib_I}(\Zsp_I)
-\frac{1}{2}\left(2U+\ell_I^2+\frac{K+m^2}{D-2}\right)
u_{a_Ib_I}(\Zsp_I)=0.
 \label{0:cEinstein-ab:Eq}
}

From Eqs.~(\ref{0:scalar eq:Eq}) and (\ref{0:cEinstein-yy:Eq}), 
we get 
\Eqrsubl{0:solution:Eq}{
&&\hspace{-0.9cm}A=A_0+\frac{1}{D-2}\ln\left[
\cos\left\{\sqrt{\frac{K+m^2}{2(D-1)}}
\left(y-y_0\right)\right\}\right],
   \label{0:A:Eq}\\
&&\hspace{-0.9cm}c_0^2=\frac{2}{(D-1)(D-2)}\,,~~~
      \label{0:c0:Eq}
}
where $A_0$ is a constant.
In terms of 
Eqs. (\ref{0:solution:Eq}), the field equations reduce to
\Eqrsubl{0:cEinstein2:Eq}{
&&\hspace{-0.5cm}R_{\mu\nu}(\Xsp)
-\frac{1}{2}\left(-f_n^2+\frac{K+m^2}{D-1}\right)q_{\mu\nu}(\Xsp)=0,
 \label{0:cEinstein2-mn:Eq}\\
&&\hspace{-0.5cm}R_{a_Ib_I}(\Zsp_I)
-\frac{1}{2}\left(\ell_I^2+\frac{K+m^2}{D-1}\right)u_{a_Ib_I}(\Zsp_I)=0\,.
 \label{0:cEinstein2-ij:Eq}
}
Hence, the metric of $D$-dimensional spacetime can be written as
\Eqr{
%\hspace{-0.3cm}
ds^2=\bar{A}^2
\left[\cos\left(\bar{y}-\bar{y}_0\right)\right]^{2/(D-2)}
\left[q_{\mu\nu}(\Xsp)dx^{\mu}dx^{\nu}
%\right.\nn\\
%&&\left.
%\hspace{-0.4cm}
+\frac{2(D-1)}{K+m^2}\,d\bar{y}^2
+\sum_Iu_{a_Ib_I}(\Zsp_I)dz^{a_I}dz^{b_I}\right],
  \label{0:D-metric3:Eq}
}
where $\bar{A}
= \e^{A_0}$ is a constant, and 
$\bar{y}$, $\bar{y}_0$ are defined by 
\Eq{
\bar{y}
= \sqrt{\frac{K+m^2}{2(D-1)}}\,y,
\quad
{\bar y}_0=
\sqrt{\frac{K+m^2}{2(D-1)}}\,y_0\,.
}
If $f_n$, $\ell_I$ and $m$ satisfy the relation 
\Eq{
\label{ds_cond}
-f_n^2+\frac{K+m^2}{D-1}>0\,,
}
the solution leads to an 
$n$-dimensional de Sitter spacetime
whose expansion rate $H$ is given by 
\Eqr{
\hspace{-0.1cm}
H^2
=\frac{1}{2(n-1)}
\left(-f_n^2+\frac{K+m^2}{D-1}\right).
}
Using Eq.~(\ref{0:cEinstein2-ij:Eq}) 
and the relation (\ref{ds_cond}), we have 
\Eq{
 R(\Zsp_I)
=\frac{L_I}{2}\left(\ell_I^2+\frac{K+m^2}{D-1}\right)
>
\frac{L_I}{2}\big(\ell_I^2+f_n^2\big)>0.
}
Hence, for the de Sitter case $R(\Xsp)>0$, 
all the internal spaces $\Zsp_I$ must be positively curved.

From Eq. (\ref{0:cEinstein2-ij:Eq}), 
the $I$th internal space can be a sphere S$^I$, if
\begin{eqnarray}
\label{int_pos}
\ell_I^2+\frac{K+m^2}{D-1}>0.
\end{eqnarray}
The $D$-dimensional metric (\ref{0:D-metric3:Eq}) 
implies that there are curvature singularities at 
$\bar{y}=\bar{y}_0+\left(n+\frac{1}{2}\right)\pi$ 
$(n~{\rm is~integer})$ because the Kretschmann 
invariant of the metric (\ref{0:D-metric3:Eq}) is given by
\Eqr{
%&&\hspace{-0.7cm}
R_{ABCD}R^{ABCD}&=&\left[
\cos\left(\bar{y}-\bar{y}_0\right)
\right]^{-\frac{4}{D-2}}\left[n(n-2)\right.\nn\\
&&\left.
+(D-n-1)(D-n-3)
+\frac{2(D-2)}{\cos^{4}\left(\bar{y}-\bar{y}_0\right)}\right].
  \label{0:Kretschmann:Eq}
}
The scalar field has the bounce configuration and diverges at 
$\bar{y}=\bar{y}_0+\left(n+\frac{1}{2}\right)\pi$, which 
gives the singularities in the $D$-dimensional background.  
Thus it is impossible to extend the spacetime across such a point
and we should restrict $\bar{y}$
to be for a  
period, e.g., 
$\bar{y}_0-\pi/2
<\bar{y}
<\bar{y}_0+\pi/2$.
Then, unless we modify the ansatz of fields and add the matter fields, 
the solutions considered in this section provide 
neither realistic cosmological models nor compactification schemes 
that lead to the finite Newton constant.

We would like to mention 
the relation of our solutions 
with the supergravity theories. 
The coupling constants of the scalar field
in the supergravity theory are severely restricted. 
Actually, coupling constants 
take the particular values, which
 are written as \cite{Minamitsuji:2010uz, Duff:1994an}
\Eqrsubl{0:cc1:Eq}{
c_n^2&=&N_n-\frac{2(n-1)(D-n-1)}{D-2}\,,
   \label{0:ccn:Eq}\\
c_I^2&=&N_I-\frac{2(L_I-1)(D-L_I-1)}{D-2}\,,
   \label{0:ccI:Eq}\\
c_0^2&=&N_0+\frac{2(D-1)}{D-2}\,,
   \label{0:cc0:Eq}
}
where $N_n$, $N_I$ and $N_0$ are constants. 
In terms of Eqs.(\ref{0:coupling:Eq}) and (\ref{0:c0:Eq}), these constants 
become 
\Eqrsubl{0:cc2:Eq}{
N_n&=&\frac{2(n-1)(D-n)}{D-1}\,,
  \label{0:nn:Eq}\\
N_I&=&\frac{2(L_I-1)(D-L_I)}{D-1}\,,
  \label{0:nI:Eq}\\
N_0&=&-\frac{2D}{D-1}\,.
  \label{0:n0:Eq}
}
Though the cases of $N_i=4,~(i=n, I, 0)$ in the ten- or eleven-dimensional 
theory correspond to supergravities, we cannot choose it due to $N_0<0$.
These supergravity solutions
always give the $\Xsp$ space to be an AdS spacetime.
In our limited ansatz of fields, it 
is difficult to obtain 
the four-dimensional de Sitter spacetime in the supergravity models.

We then discuss the case of $m=0$. 
We assume that the $D$-dimensional metric and the gauge field strengths 
take the same form as (\ref{0:metric:Eq}), (\ref{0:ansatz of gauge1:Eq}) 
and (\ref{0:ansatz of gauge2:Eq}). Then, the Einstein equations become 
\Eqrsubl{v0:cEinstein0:Eq}{
&&\hspace{-1.1cm}
R_{\mu\nu}(\Xsp)-\left[U
-\frac{1}{2(D-2)}\left\{(D-n-3)f_n^2\,\e^{-2(n-1)A
+\varepsilon c_n\phi}\right.\right.\nn\\
&&\left.\left.~~+\sum_I(L_I-1)\ell_I^2\,\e^{-2(L_I-1)A
+\varepsilon c_I\phi}\right\}\right]q_{\mu\nu}(\Xsp)
=0,
 \label{v0:cEinstein-mn0:Eq}\\
&&\hspace{-1.1cm}
-(D-1)\pd_y^2A-\frac{2}{c_0^2}\left(\pd_yA\right)^2
-\frac{1}{2(D-2)}\left[(n-1)f_n^2\,\e^{-2(n-1)A
+\varepsilon c_n\phi}\right.\nn\\
&&\left.~~-\sum_I(L_I-1)\ell_I^2\,\e^{-2(L_I-1)A
+\varepsilon c_I\phi}\right]
=0,
  \label{v0:cEinstein-yy0:Eq}\\
&&\hspace{-1.1cm}R_{a_Ib_I}(\Zsp_I)
-\left[U+\frac{1}{2(D-2)}\left\{(n-1)f_n^2\,\e^{-2(n-1)A
+\varepsilon c_n\phi}\right.\right.\nn\\
&&\left.\left.~~+\sum_I(D-L_I-3)\ell_I^2\,\e^{-2(L_I-1)A
+\varepsilon c_I\phi}\right\}
\right]u_{a_Ib_I}(\Zsp_I)=0,
 \label{v0:cEinstein-ab0:Eq}
}
where $R_{\mu\nu}(\Xsp)$, $R_{a_Ib_I}(\Zsp_I)$ are the Ricci tensors
of the metrics $q_{\mu\nu}(\Xsp)$, $u_{a_Ib_I}(\Zsp_I)$, and 
$U$ is defined by \eqref{0:U:Eq}. 
Let us next consider the scalar field. 
For coupling constants of the scalar field, 
we choose 
\Eq{
-2(n-1)A+\varepsilon c_n\phi=0\,,~~~~-2(L_I-1)A
+\varepsilon c_I\phi=0.
  \label{v0:ansatz of scalar:Eq}
}
Among these equations, the first together with the second assumption 
in \eqref{v0:ansatz of scalar:Eq} reads  
\begin{equation}
c_{I}=\left(n-1\right)^{-1}\left(L_I-1\right)c_n\,.
   \label{v0:coupling:Eq}
\end{equation} 
Upon setting the assumptions of fields and 
the coupling constants (\ref{v0:coupling:Eq}), 
the Bianchi identities and 
the equations of motion for the
gauge fields are again automatically satisfied. 
In terms of the assumptions (\ref{0:metric:Eq}) and 
the field 
equations  
reduce to
Eqs. (\ref{0:cEinstein2:Eq}) and (\ref{0:D-metric3:Eq})
with $m=0$. 
The \eqref{0:cEinstein2-mn:Eq} with $m=0$ 
illustrates that de Sitter compactification is not allowed
(see \cite{Maldacena:2000mw} for the general cases). 
On the other hand, for $m\ne 0$, the $\Xsp$ space 
can be an Einstein space with a positive curvature 
where the existence of de Sitter solutions 
is clear. 
Each internal space $\Zsp_I$
can be sphere S$^I$, if \eqref{int_pos} is satisfied.  

In the case of $m=0$, there are solutions of supergravity in ten dimensions.
From (\ref{0:cc1:Eq}) and (\ref{0:cc2:Eq}), 
the solution of $(D, n, L_1, L_2)=(10, 4, 1, 4)$
with $m=0$ and $\ell_1=0$
(here $I=1,2$) gives
that of ten-dimensional type 
IIA supergravity.
Similarly,
the solution of
$(D, n, L_1)=(10, 2, 7)$ with $f_2=0$ and 
$m=0$
(here $I=1$)
gives that
of ten-dimensional type IIB supergravity.
However, these solutions only 
lead to AdS compactifications.

%======================================%
%<<<<<<<<<<<<< SECTION 3 >>>>>>>>>>>>>>%
%======================================%
%T1>Time dependent internal space
\section{Time-dependent internal space}
\label{sec:s}

In this section, 
we investigate
an embedding of an expanding FLRW universe
into a time-dependent warped spacetime.

Let us consider a gravitational theory with the metric $g_{MN}$,
the scalar field $\phi$, the cosmological constant $\Lambda$,
and the antisymmetric tensor field of rank $p_I$.
The action which we consider is given by
\Eq{
S=\frac{1}{2\kappa^2}\int 
\left[\left(R-2\e^{\alpha\phi}\Lambda\right)\ast{\bf 1}
-\frac{1}{2}d\phi\wedge\ast d\phi
-\sum_I\frac{1}{2\cdot p_I!}\e^{\alpha_I\phi}
F_{(p_I)}\wedge\ast F_{(p_I)}\right],
\label{t:action:Eq}
}
where $\kappa^2$ is the $D$-dimensional gravitational constant, and 
$\ast$ is the Hodge operator in the $D$-dimensional spacetime, 
$F_{(p_I)}$ is the $p_I$-form field strength,
and $\alpha$, $\alpha_I$ are constants.

The $D$-dimensional action \eqref{t:action:Eq} gives the 
field equations: 
\Eqrsubl{t:field equations:Eq}{
&&\hspace{-0.9cm}R_{MN}=\frac{1}{2}\pd_M\phi\pd_N\phi
+\frac{2}{D-2}\e^{\alpha\phi}\Lambda g_{MN}\nn\\
&&~~~~~+\sum_I\frac{\e^{\alpha_I\phi}}{2\cdot p_{I}!}
\left[p_I\left(F_{(p_I)}\right)^2_{MN}
-\frac{p_I-1}{D-2}g_{MN}F_{(p_I)}^2\right],
   \label{t:Einstein:Eq}\\
&&\hspace{-0.9cm}
d\left[\e^{\alpha_I\phi}\,\ast F_{(p_I)}\right]=0\,,
    \label{t:gauge:Eq}\\
&&\hspace{-0.9cm}d\ast d\phi-2\alpha\e^{\alpha\phi}\Lambda\ast{\bf 1}
-\sum_I\frac{\alpha_I}{2\cdot p_I!}\e^{\alpha_I\phi}F_{(p_I)}
\wedge\ast F_{(p_I)}=0\,,
   \label{t:scalar:Eq}
}
where 
$\left(F_{(p_I)}\right)^2_{MN}
= F_{MA_1\cdots A_{p_I-1}}
{F_N}^{A_1\cdots A_{p_I-1}}$.
To solve the field equations, we assume that the $D$-dimensional metric
takes the form
\Eq{
ds^2=\e^{2\left[A_0(x)+A_1(y)\right]}\left[q_{\mu\nu}(\Xsp)dx^{\mu}dx^{\nu}
+dy^2(\Ysp)+\sum_Iu_{a_Ib_I}(\Zsp_I)dz^{a_I}dz^{b_I}\right],
 \label{t:metric:Eq}
}
where $q_{\mu\nu}(\Xsp)$ is  
the $n$-dimensional metric which
depends only on the $n$-dimensional coordinates $x^{\mu}$,
and $u_{a_Ib_I}(\Zsp_I)$ is the $L_I$-dimensional metric which
depends only on the $L_I$-dimensional coordinates $z^{a_I}$, 
the functions 
$A_0(x)$ and $A_1(y)$ depend only on the coordinates 
$x^{\mu}$ and $y$,  respectively.

Furthermore, we consider the field strength $F_{(L_I)}$.  
The scalar field $\phi$ and
the gauge field strengths are assumed to be
\Eqrsubl{t:fields:Eq}{
\phi&=&-\frac{2}{\alpha} \left[A_0(x)+A_1(y)\right]\,,
  \label{t:ansatz of scalar:Eq}\\
F_{(L_I)}&=&f_I\,\omega_I\,,
  \label{t:ansatz of gauge:Eq}
}
where $f_I$ is a constant, and 
$\omega_I$ denotes the volume $L_I$-form of the internal space 
$\Zsp_I$.
In the following, we assume that the parameter $\alpha_I$ 
is given by 
\Eq{
\alpha_I=-\alpha\left(L_I-1\right)\,.
   \label{t:coupling:Eq}
}

Let us first consider the Einstein Eqs.~(\ref{t:Einstein:Eq}).
Using the assumptions (\ref{t:metric:Eq}) and (\ref{t:fields:Eq}),
the Einstein equations are given by
\Eqrsubl{t:cEinstein:Eq}{
&&\hspace{-1.1cm}R_{\mu\nu}(\Xsp)-(D-2)D_{\mu}D_{\nu}A_0
+\left(D-2-\frac{2}{\alpha^2}\right)\pd_{\mu}A_0\pd_{\nu}A_0
-\left(\bar{U}+\bar{K}\right)q_{\mu\nu}(\Xsp)=0,
 \label{t:cEinstein-mn:Eq}\\
&&\hspace{-1.1cm}\left(D-2-\frac{2}{\alpha^2}\right)\pd_{\mu}A_0\pd_yA_1=0\,,
 \label{t:cEinstein-my:Eq}\\
&&\hspace{-1.1cm}(D-2)\pd_y^2A_1
-\left(D-2-\frac{2}{\alpha^2}\right)\left(\pd_yA_1\right)^2
+\bar{U}+\bar{K}=0\,,
 \label{t:cEinstein-yy:Eq}\\
&&\hspace{-1.1cm}R_{a_Ib_I}(\Zsp_I)
-\left(\frac{f_I^2}{2}+\bar{U}+\bar{K}\right)u_{a_Ib_I}(\Zsp_I)=0\,,
 \label{t:cEinstein-ij:Eq}
}
where $D_{\mu}$ is the covariant derivative with respect to
the metric $q_{\mu\nu}$ and $\triangle_{\Xsp}$ is  
the Laplace operator on the space of X, and
$R_{\mu\nu}(\Xsp)$, $R_{a_Ib_I}(\Zsp_I)$ are the Ricci tensors
of the metrics $q_{\mu\nu}$, $u_{a_Ib_I}$, respectively, and 
$\bar{U}$, $\bar{K}$ are defined as
\Eqrsubl{t:parameter:Eq}{
\bar{U}&=&\lap_{\Xsp}A_0+(D-2)q^{\rho\sigma}\pd_{\rho}A_0
\pd_{\sigma}A_0+\pd_y^2A+(D-2)\left(\pd_yA\right)^2\,,\\
\bar{K}&=& \frac{1}{2(D-2)}\left[4\Lambda-\sum_I(L_I-1)f_I^2\right]\,.
  \label{t:K:Eq}
} 
From \eqref{t:cEinstein-my:Eq}, the parameter $\alpha$ must be
in the form
\Eq{
\alpha=\pm\sqrt{\frac{2}{D-2}}\,.
   \label{t:alpha:Eq}
}

Next we consider the gauge fields. 
Under the assumption \eqref{t:ansatz of gauge:Eq}, 
the Bianchi identities and 
the equations of motion for the
gauge fields are automatically satisfied. 
Substituting Eqs.~(\ref{t:metric:Eq}), (\ref{t:fields:Eq}) and 
(\ref{t:alpha:Eq}) into
Eq.~(\ref{t:scalar:Eq}), the scalar field equation gives
\Eqr{
\bar{U}+\bar{K}=0\,.
   \label{t:scalar eq:Eq}
}
In terms of \eqref{t:fields:Eq} and \eqref{t:alpha:Eq}, 
the field equations reduce to
\Eqrsubl{t:cEinstein2:Eq}{
&&\hspace{-0.5cm}R_{\mu\nu}(\Xsp)-(D-2)D_{\mu}D_{\nu}A_0=0,
 \label{t:cEinstein2-mn:Eq}\\
&&\hspace{-0.5cm}(D-2)\pd_y^2A_1=0\,,
 \label{t:cEinstein2-yy:Eq}\\
&&\hspace{-0.5cm}R_{a_Ib_I}(\Zsp_I)-\frac{f_I^2}{2}\,u_{a_Ib_I}(\Zsp_I)=0,
 \label{t:cEinstein2-ij:Eq}\\
&&\hspace{-0.5cm}\lap_{\Xsp}A_0+(D-2)q^{\rho\sigma}\pd_{\rho}A_0
\pd_{\sigma}A_0+\pd_y^2A+(D-2)\left(\pd_yA\right)^2
\nn\\
&&+\frac{2}{2(D-2)}\left[4\Lambda-\sum_I(L_I-1)f_I^2\right]=0\,.
   \label{t:scalar eq2:Eq}
}
Note that the \eqref{t:cEinstein2-mn:Eq} is the differential 
equation with respect to the coordinates $x^{\mu}$ 
while Eqs.~(\ref{t:cEinstein2-yy:Eq}) 
are the equations of $y$. 
Thus, we can treat them separately. 
We also find that
from Eq. (\ref{t:cEinstein2-ij:Eq}),
each internal space $\Zsp_I$
is positively curved if 
the corresponding
field strength is nonvanishing
along $\Zsp_I$.

Let us first consider the \eqref{t:cEinstein2-yy:Eq}. 
The solution of $A_1$ is 
\Eq{
A_1(y)=\ell(y-y_0)\,,
}
where $\ell$ is a constant. 

Next we consider the \eqref{t:cEinstein2-mn:Eq}. 
We set the function $A_0$ and 
the $n$-dimensional spacetime metric 
$q_{\mu\nu}$ to be a spatially homogeneous function and 
a spatially flat 
FLRW universe, respectively. 
With the scale factor $a$ turned on, 
it turns out that the metric is of the form 
\Eqrsubl{t:metric-n:Eq}{
A_0&=&A_0(t)\,,
    \label{t:a0:Eq}\\
q_{\mu\nu}(\Xsp)dx^{\mu}dx^{\nu}
  &=&-dt^2+a^2(t)\delta_{mn}dx^{m}dx^{n}\,,
    \label{t:n-metric:Eq}
}
where $\delta_{
mn}$ is the metric of $(n-1)$-dimensional Euclidean space. 

Using the $n$-dimensional metric (\ref{t:metric-n:Eq}),  
the Eq.~(\ref{t:cEinstein-mn:Eq}) is rewritten by
\Eqrsubl{t:cEinstein3:Eq}{
&&\hspace{-1.2cm}(n-1)\left[\left(\frac{\dot{a}}{a}\right)^{\dot{}}
+\left(\frac{\dot{a}}{a}\right)^2\right]+(D-2)\ddot{A}_0=0,
    \label{t:cEinstein3-tt:Eq}\\
&&\hspace{-1.2cm}
\left(\frac{\dot{a}}{a}\right)^{\dot{}}
+(n-1)\left(\frac{\dot{a}}{a}\right)^2
+(D-2)\frac{\dot{a}}{a}\,\dot{A}_0
=0,
   \label{t:cEinstein3-ab:Eq}
}
where %$\dot{}$
``dot'' denotes the ordinary derivative 
with respect to the coordinate $t$.
We assume that 
the functions $a(t)$ and $A_0(t)$ 
are given by 
\Eq{
\frac{\dot{a}}{a}=c_1\,t^{-1}\,,~~~~~~\dot{A}_0=c_2\,t^{-1}\,,
  \label{t:ansatz:Eq}
}
where $c_1$ and $c_2$ are constants.
Thus, the Einstein equations (\ref{t:cEinstein3:Eq}) reduce to 
\Eqrsubl{t:cEinstein4:Eq}{
&&(n-1)c_1\left(c_1-1\right)-(D-2)c_2=0\,,
    \label{t:cEinstein4-tt:Eq}\\
&&\left[-1+(n-1)c_1+(D-2)c_2\right]c_1=0\,.   
    \label{t:cEinstein4-mn1:Eq}
}
Except for the trivial solution $c_1=c_2=0$\,, 
we find  
\Eqr{
c_1=\pm\left(n-1\right)^{-1/2}\,,~~~~~
c_2=\frac{
\mp (n-1)+\sqrt{n-1}}{(D-2)\sqrt{n-1}}\,.
\label{bra}
}
From the Eq.~(\ref{t:ansatz:Eq}), 
we have 
\Eq{
a(t)=a_0\,\left(t-t_0\right)^{c_1}\,,~~~~
A_0(t)=c_2\ln\left[\psi_0\left(t-t_0\right)\right]\,,
   \label{t:sol for Aa:Eq}
}
where $\psi_0$ and $a_0$ are constants. 
Then the metric of the $D$-dimensional spacetime can be written as
\Eqr{
ds^2=\e^{2A_1(y)}\left[-d\tau^2
+a_0^2\,\left(\frac{\tau}{\tau_0}\right)^{\frac{2(c_1+c_2)}{c_2+1}}
\delta_{mn}dx^{m}dx^{n}
%\right.\nn\\
%&&\left. 
+\left(\frac{\tau}{\tau_0}\right)^{\frac{2c_2}{c_2+1}}
\left(dy^2+u_{ab}(\Zsp)dz^adz^b\right)\right],
  \label{t:D-metric2:Eq}
}
where the cosmic time $\tau$ and the parameter $\tau_0$ are
given by 
\Eq{
\frac{\tau}{\tau_0}=t^{c_2+1}\,,
~~~~~\tau_0=\frac{1}{c_2+1}\,.
\label{t:ctime:Eq}
}
We find 
power-law cosmological evolutions. 
The power of the expansion of the external
$(n-1)$-dimensional space is given by
\Eqr{
\lambda=\frac{c_1+c_2}{c_2+1}\,.
   \label{t:power:Eq}
}
For $D=10$ and $n=4$,  
we find that the fastest expanding case has the power 
$\lambda=0.53478$ 
in terms of the proper time $\tau$,
while the lower branch one 
gives 
$\lambda=-0.175805$.
Since we have obtained $\lambda<1$ 
for $n=4$, these
solutions do not lead to accelerated expansion. 
However, by taking $\tau=\bar{\tau_{\rm c}}-\bar{\tau}$, 
we get
\Eqr{
ds^2&=&\e^{2A_1(y)}\left[-d\bar{\tau}^2
+a_0^2\,\left(\frac{\bar{\tau}_{\rm c}-\bar{\tau}}
{\tau_0}\right)^{\frac{2(c_1+c_2)}{c_2+1}}
\delta_{mn}dx^{m}dx^{n}
\right.\nn\\
&&\left. 
+\left(\frac{\bar{\tau}_{\rm c}-\bar{\tau}}{\tau_0}\right)^{\frac{2c_2}{c_2+1}}
\left(dy^2+u_{ab}(\Zsp)dz^adz^b\right)\right],
  \label{t:D-metric3:Eq}
}
where $\bar{\tau}_{\rm c}$ is a constant. 
For $\bar{\tau}<\bar{\tau}_{\rm c}$ we have accelerated
expansion. Since $q_{\mu\nu}(\Xsp)$ is not the Einstein-frame metric, 
we consider the cosmic expansion in the Einstein frame. 
In order to analyze in the Einstein
frame, therefore we perform the following conformal transformation
to make the non-Einstein conformal frame into the Einstein frame:
\Eq{
q_{\mu\nu}(\Xsp)=\left(\frac{\tau}{\tau_0}\right)^{\frac{-4c_2(D-n)}{(c_2+1)(n-2)}}
q_{\mu\nu}(\bXsp)\,,
}
where $q_{\mu\nu}(\bXsp)$ is the $n$-dimensional metric 
in the Einstein frame. 
Then, the metric of $n$-dimensional spacetime is given by 
\Eq{
q_{\mu\nu}(\bXsp)dx^{\mu}dx^{\nu}=-d\tilde{\tau}^2+
\left(\frac{\tilde{\tau}}
{\tilde{\tau}_0}\right)^{2\frac{c_1(n-2)+c_2(2D-n-2)}{c_2(2D-n-2)+n-2}}
\delta_{mn}dx^{m}dx^{n}\,,
}
where $\tilde{\tau}_0$ is a constant and $\tilde{\tau}$ is defined by
\Eq{
\frac{\tilde{\tau}}{\tilde{\tau}_0}
=\left(\frac{\tau}{\tau_0}\right)^{1+\frac{2c_2(D-n)}{(c_2+1)(n-2)}}.
} 
The power of time-dependence in 
the scale factor is given by
\Eq{
0<\frac{c_1+c_2(D-3)}{1+c_2(D-3)}<1, ~~~{\rm For}~~n=4,~D>5\,.
%0<\frac{c_1(n-2)+c_2(2D-n-2)}{c_2(2D-n-2)+n-2}<1, ~~~{\rm For}~~n=4,~D>5\,.
}
Hence, the Einstein frame metric $q(\bXsp)$ yields the decelerating
universe.

%======================================%
%<<<<<<<<<<<<< SECTION 4 >>>>>>>>>>>>>>%
%======================================%
%T1>Discussions
\section{Discussions}
 \label{sec:Discussions}
In this work, 
we have presented exact solutions of 
the warped compactification 
on S${}^1\times \prod_I$S$^{L_I}$,
as in  
our recent study  
on the warped  
compactification on S$^1\times$S$^{D-n-1}$ 
\cite{Minamitsuji:2011xs}. 

We gave a new class of warped de Sitter solutions 
and illustrated how the matter fields,
in particular the 0-form field strength, 
provide a de Sitter spacetime. 
The Einstein equations have led to a de Sitter spacetime 
in the presence of the 0-form field strength and constant fluxes.
If there is no 0-form, 
the Einstein equations require AdS compactifications.
We note that it is easy to find a solution if flux terms are not 
allowed to depend on the extra dimensional coordinate $y$. 
This solution is different
from solutions found in \cite{Minamitsuji:2011xs}. 
However, the assumptions (\ref{0:fields:Eq}) and (\ref{0:coupling:Eq})  
may play an important role in constructing a de Sitter solution. 
We also presented warped cosmological solutions 
where the warp factor is time-dependent. 
Although we could find solutions in which 
the time-dependent warp factor allows the accelerated expansion
in the non-Einstein conformal frame in the higher-dimensional spacetime,
the solutions presented here could not give 
accelerated expansion in the Einstein frame. 
However, it would be very interesting to explore such solutions 
and to see whether and how the time-dependence with the warped 
structure of spacetime could realize accelerated expansion. 

Our result for the cosmological solutions consists 
of two pieces.
The first is a singular solution with a spherical 
internal space; the second is 
a nonsingular solution with 
an infinite volume internal space.
The solution which we desire should 
have a compact internal space,
where the conventional dimensional reduction scheme 
can be applied.
Although the examples illustrated in 
this paper provide neither realistic cosmological
models nor compactification schemes, 
the feature of the solutions or
the method to obtain them could open new directions to study 
how to construct accelerated expansion of the universe 
as well as an appropriate higher-dimensional cosmological model.

%T1>Acknowledgments
\section*{Acknowledgments}
The work of M.M. was supported by Yukawa fellowship and 
by Grant-in-Aid for Young Scientists (B) of JSPS Research,
under Contract No. 24740162. 
K.U. was supported by Grant-in-Aid for 
Young Scientists (B) of JSPS Research, under Contract No. 20740147.

%======================================%
%<<<<<<<<<<<<< REFERENCE >>>>>>>>>>>>>>%
%======================================%

%T1>References
%\section*{References}
%\begin{thebibliography}{10}


\begin{thebibliography}{99}
%\cite{Tsujikawa:2010sc}
\bibitem{Tsujikawa:2010sc}
  S.~Tsujikawa,
  ``Dark energy: investigation and modeling,''
  arXiv:1004.1493 [astro-ph.CO].

%\cite{Minamitsuji:2011xs}
\bibitem{Minamitsuji:2011xs}
  M.~Minamitsuji and K.~Uzawa,
  ``Warped de Sitter compactifications in the scalar-tensor theory,''
  Phys.\ Lett.\  B {\bf 710} (2012) 358
  [arXiv:1109.4818 [hep-th]].

%\cite{Neupane:2009jn}
\bibitem{Neupane:2009jn}
  I.~P.~Neupane,
  ``Simple cosmological de Sitter solutions on 
  dS$_4 \times {\rm Y}_6$ spaces,''
  Class.\ Quant.\ Grav.\  {\bf 27} (2010) 045011
  [arXiv:0901.2568 [hep-th]].

%\cite{Neupane:2010is}
\bibitem{Neupane:2010is}
  I.~P.~Neupane,
  ``Warped compactification on curved manifolds,''
  Class.\ Quant.\ Grav.\  {\bf 28} (2011) 125015
  [arXiv:1006.4495 [hep-th]].

%\cite{Neupane:2010ya}
\bibitem{Neupane:2010ya}
  I.~P.~Neupane,
  ``Warped compactification to de Sitter space,''
  Nucl.\ Phys.\  B {\bf 847} (2011) 549
  [arXiv:1011.5007 [hep-th]].

%\cite{Minamitsuji:2011gn}
\bibitem{Minamitsuji:2011gn}
  M.~Minamitsuji and K.~Uzawa,
  ``Spectrum from the warped compactifications with the de Sitter universe,''
  JHEP {\bf 1207} (2012) 154
  [arXiv:1103.5325 [hep-th]].

%\cite{Minamitsuji:2011gp}
\bibitem{Minamitsuji:2011gp}
  M.~Minamitsuji and K.~Uzawa,
  ``Warped de Sitter compactifications,''
  JHEP {\bf 1201} (2012) 142
  [arXiv:1103.5326 [hep-th]].

%\cite{Lu:1996jk}
\bibitem{Lu:1996jk}
  H.~Lu, S.~Mukherji, C.~N.~Pope and K.~W.~Xu,
  ``Cosmological solutions in string theories,''
  Phys.\ Rev.\  D {\bf 55} (1997) 7926
  [arXiv:hep-th/9610107].

%\cite{Behrndt:2003cx}
\bibitem{Behrndt:2003cx}
  K.~Behrndt and M.~Cvetic,
  ``Time-dependent backgrounds from supergravity with gauged non-compact
  $R$-symmetry,''
  Class.\ Quant.\ Grav.\  {\bf 20} (2003) 4177
  [arXiv:hep-th/0303266].

%\cite{Gibbons:2005rt}
\bibitem{Gibbons:2005rt}
  G.~W.~Gibbons, H.~Lu and C.~N.~Pope,
  ``Brane worlds in collision,''
  Phys.\ Rev.\ Lett.\  {\bf 94} (2005) 131602
  [arXiv:hep-th/0501117].
  
%\cite{Chen:2005jp}
\bibitem{Chen:2005jp}
  W.~Chen, Z.~W.~Chong, G.~W.~Gibbons, H.~Lu and C.~N.~Pope,
  ``Horava-Witten stability: Eppur si muove,''
  Nucl.\ Phys.\  B {\bf 732} (2006) 118
  [arXiv:hep-th/0502077].

%\cite{Kodama:2005fz}
\bibitem{Kodama:2005fz}
  H.~Kodama and K.~Uzawa,
  ``Moduli instability in warped compactifications of the type IIB
  supergravity,''
  JHEP {\bf 0507} (2005) 061
  [arXiv:hep-th/0504193].

%\cite{Kodama:2005cz}
\bibitem{Kodama:2005cz}
  H.~Kodama and K.~Uzawa,
  ``Comments on the four-dimensional effective theory for warped
  compactification,''
  JHEP {\bf 0603} (2006) 053
  [arXiv:hep-th/0512104].

%\cite{Kodama:2006ay}
\bibitem{Kodama:2006ay}
  H.~Kodama and K.~Uzawa,
  ``Moduli instability in warped compactification,''
  arXiv:hep-th/0601100.

%\cite{Binetruy:2007tu}
\bibitem{Binetruy:2007tu}
  P.~Binetruy, M.~Sasaki and K.~Uzawa,
  ``Dynamical D4-D8 and D3-D7 branes in supergravity,''
  Phys.\ Rev.\  D {\bf 80} (2009) 026001
  [arXiv:0712.3615 [hep-th]].

%\cite{Binetruy:2008ev}
\bibitem{Binetruy:2008ev}
  P.~Binetruy, M.~Sasaki and K.~Uzawa,
  ``Dynamical solution of supergravity,''
  arXiv:0801.3507 [hep-th].

%\cite{Maeda:2009tq}
\bibitem{Maeda:2009tq}
  K.~i.~Maeda, N.~Ohta, M.~Tanabe and R.~Wakebe,
  ``Supersymmetric Intersecting Branes in Time-dependent Backgrounds,''
  JHEP {\bf 0906} (2009) 036
  [arXiv:0903.3298 [hep-th]].

%\cite{Maeda:2009zi}
\bibitem{Maeda:2009zi}
  K.~i.~Maeda, N.~Ohta and K.~Uzawa,
  ``Dynamics of intersecting brane systems -- Classification and their
  applications --,''
  JHEP {\bf 0906} (2009) 051
  [arXiv:0903.5483 [hep-th]].

%\cite{Gibbons:2009dr}
\bibitem{Gibbons:2009dr}
  G.~W.~Gibbons and K.~i.~Maeda,
  ``Black Holes in an Expanding Universe,''
  Phys.\ Rev.\ Lett.\  {\bf 104} (2010) 131101
  [arXiv:0912.2809 [gr-qc]].

%\cite{Maeda:2009ds}
\bibitem{Maeda:2009ds}
  K.~i.~Maeda and M.~Nozawa,
  ``Black Hole in the Expanding Universe from Intersecting Branes,''
  Phys.\ Rev.\  D {\bf 81} (2010) 044017
  [arXiv:0912.2811 [hep-th]].

%\cite{Uzawa:2010zza}
\bibitem{Uzawa:2010zza}
  K.~Uzawa,
  ``Dynamical intersecting brane solutions of supergravity,''
  AIP Conf.\ Proc.\  {\bf 1200} (2010) 541.
  
%\cite{Maeda:2010yk}
\bibitem{Maeda:2010yk}
  K.~i.~Maeda, N.~Ohta, M.~Tanabe and R.~Wakebe,
  ``Supersymmetric Intersecting Branes on the Waves,''
  JHEP {\bf 1004} (2010) 013
  [arXiv:1001.2640 [hep-th]].

%\cite{Maeda:2010ja}
\bibitem{Maeda:2010ja}
  K.~i.~Maeda and M.~Nozawa,
  ``Black Hole in the Expanding Universe with Arbitrary Power-Law Expansion,''
  Phys.\ Rev.\  D {\bf 81} (2010) 124038
  [arXiv:1003.2849 [gr-qc]].

%\cite{Minamitsuji:2010fp}
\bibitem{Minamitsuji:2010fp}
  M.~Minamitsuji, N.~Ohta and K.~Uzawa,
  ``Dynamical solutions in the 3-Form Field Background in the
  Nishino-Salam-Sezgin Model,''
  Phys.\ Rev.\  D {\bf 81} (2010) 126005
  [arXiv:1003.5967 [hep-th]].

%\cite{Maeda:2010aj}
\bibitem{Maeda:2010aj}
  K.~i.~Maeda, M.~Minamitsuji, N.~Ohta and K.~Uzawa,
  ``Dynamical $p$-branes with a cosmological constant,''
  Phys.\ Rev.\  D {\bf 82} (2010) 046007
  [arXiv:1006.2306 [hep-th]].

%\cite{Minamitsuji:2010kb}
\bibitem{Minamitsuji:2010kb}
  M.~Minamitsuji, N.~Ohta and K.~Uzawa,
  ``Cosmological intersecting brane solutions,''
  Phys.\ Rev.\  D {\bf 82} (2010) 086002
  [arXiv:1007.1762 [hep-th]].

%\cite{Minamitsuji:2010uz}
\bibitem{Minamitsuji:2010uz}
  M.~Minamitsuji and K.~Uzawa,
  ``Cosmology in $p$-brane systems,''
  Phys.\ Rev.\  D {\bf 83} (2011) 086002
  [arXiv:1011.2376 [hep-th]].
  
%\cite{Uzawa:2010zz}
\bibitem{Uzawa:2010zz}
  K.~Uzawa,
  ``Cosmological intersecting brane solutions in string theory,''
  J.\ Phys.\ Conf.\ Ser.\  {\bf 259} (2010) 012032.

%\cite{Maeda:2011sh}
\bibitem{Maeda:2011sh}
  K.~i.~Maeda and M.~Nozawa,
  ``Black hole solutions in string theory,''
  Prog.\ Theor.\ Phys.\ Suppl.\  {\bf 189} (2011) 310
  [arXiv:1104.1849 [hep-th]].

%\cite{Minamitsuji:2011jt}
\bibitem{Minamitsuji:2011jt}
  M.~Minamitsuji and K.~Uzawa,
  ``Dynamics of partially localized brane systems,''
  Phys.\ Rev.\  D {\bf 84} (2011) 126006
  [arXiv:1109.1415 [hep-th]].

%\cite{Maeda:2012xb}
\bibitem{Maeda:2012xb}
  K.~i.~Maeda and K.~Uzawa,
  ``Dynamical brane with angles : Collision of the universes,''
  Phys.\ Rev.\  D {\bf 85} (2012) 086004
  [arXiv:1201.3213 [hep-th]].

%\cite{Minamitsuji:2012if}
\bibitem{Minamitsuji:2012if}
  M.~Minamitsuji and K.~Uzawa,
  ``Cosmological brane systems in warped spacetime,''
  Phys.\  Rev.\ D {\bf 87} (2013) 046010
  [arXiv:1207.4334 [hep-th]].

%\cite{Duff:1994an}
\bibitem{Duff:1994an}
  M.~J.~Duff, R.~R.~Khuri and J.~X.~Lu,
  ``String solitons,''
  Phys.\ Rept.\  {\bf 259} (1995) 213
  [arXiv:hep-th/9412184].

%\cite{Maldacena:2000mw}
\bibitem{Maldacena:2000mw} 
  J.~M.~Maldacena and C.~Nunez,
  ``Supergravity description of field theories on curved manifolds and a no go theorem,''
  Int.\ J.\ Mod.\ Phys.\ A {\bf 16}, 822 (2001)
  [hep-th/0007018].

\end{thebibliography}
\end{document}